\documentstyle[preprint,aps]{revtex}

\begin{document}

\newcommand{\beq}{\begin{equation}}
\newcommand{\eeq}{\end{equation}}
\newcommand{\bea}{\begin{eqnarray}}
\newcommand{\eea}{\end{eqnarray}}
\newcommand{\rv}{\rangle}
\newcommand{\lv}{\langle}

\draft
\tighten
\preprint{\vbox{
\hbox{IC-98-33}
\hbox{IJS-TP-98-4}
\hbox{CERN-TH/98-3}
}}

\title{R-charge Kills Monopoles}

\author{ Borut Bajc \thanks{ borut.bajc@ijs.si}}
\address{International Center for Theoretical Physics,
34100 Trieste, Italy, {\rm and} \newline J. Stefan 
Institute, 1001  Ljubljana, Slovenia   } 
\author{ Antonio Riotto \thanks{riotto@nxth04.cern.ch }}
\address{CERN, Theory Group, CH-1211, Geneva 23, Switzerland}
\author{ Goran Senjanovi\'c \thanks{ 
goran@ictp.trieste.it}}
\address{International Center for Theoretical Physics,
34100 Trieste, Italy }

\maketitle

\begin{abstract}

Large charge density, unlike high temperature, may  lead to 
nonrestoration of global and gauge symmetries. 
Supersymmetric GUTs with the appealing scenario of unification
scale  being generated dynamically naturally contain global 
continuous $R$ symmetries. We point out that the presence of a large 
$R$ charge in the early Universe can lead to GUT symmetry
nonrestoration. This provides a simple way out of the monopole
problem.
 
\end{abstract}

\newpage

\section{Introduction}
\label{intr}

The existence of magnetic monopoles is one of the most 
beautiful aspects  of the idea of Grand Unification. 
Unfortunately, due to their super large mass and 
overproduction in the early Universe, this at the same 
time represents a cosmological catastrophe known as the 
monopole problem \cite{p79}. The conventional solution 
to this problem could be divided into three categories: 
inflation \cite{g81}, Langacker-Pi mechanism \cite{lp80} 
and symmetry nonrestoration \cite{dms95}. Recently, 
another interesting scenario has been suggested in which 
the monopoles are swept away by domain walls \cite{dlv97}. 

In this paper we focus on the symmetry nonrestoration scenario, 
which in itself is a fascinating phenomenon that defies common 
intuition \cite{w74,ms79}. Unfortunately, it may not work in gauge 
theories due to the large next to leading order 
correction \cite{bl96}. On the other hand it has been 
known that a large background charge density provides a 
natural setting for the breakdown of gauge symmetries in 
the early Universe \cite{l76,l79,ls94}. We showed recently 
that this may provide a simple solution of the monopole 
problem based on the simple extension of the standard 
model and a large lepton number \cite{brs97}. 

In recent years supersymmetric Grand Unified Theories (GUTs) 
have become increasingly more popular for two fundamental 
reasons. Low energy supersymmetry naturally protects large 
mass hierarchies and, equally 
important, it leads to the unification of gauge 
couplings \cite{ej82}. It is thus a particular 
challenge to solve the monopole problem in supersymmetric 
GUTs. In this context a large background charge may be 
important for it has been shown recently that it provides 
high temperature symmetry nonrestoration in supersymmetry 
too \cite{rs97}. The point is that without any external charge, 
in supersymmetry the internal symmetries are necessarily restored 
\cite{h82,m84}. This is true even when one includes the higher 
dimensional operators \cite{bms96}, in spite of some interesting 
attempts on the contrary \cite{dt96}.

In this letter we point out that the 
role of the background charge of the Universe may be 
naturally played by global $R$ charges. Our motivation and 
inspiration lies in the simple well known fact that the 
Minimal Supersymmetric Standard Model (MSSM) possesses a 
global $U(1)_R$ symmetry. 
Actually, this is true even in the general case when all 
the $R$-parity breaking terms are allowed \cite{ds94}. Of 
course, the soft supersymmetry breaking terms in the 
potential also break this $U(1)_R$ symmetry, so that 
today the background $R$ charge of the Universe would have 
necessarily been washed out. However, at very high 
temperature their effects get suppressed \cite{rs97} so 
that it is perfectly sensible to speak of possibly large 
and conserved $R$ charge in the early Universe. This is not 
the whole story though, since this $R$ charge must also be 
compatible with Grand Unification. Our work is devoted 
precisely to this issue. In the following sections we 
argue that $U(1)_R$ symmetries are naturally present in 
supersymmetric GUTs, which generate large mass 
hierarchies dynamically. In such theories there may be no 
monopole problem whatsoever.

In what follows we first give a simple example of a gauge 
model with an automatic global $U(1)_R$ symmetry and 
discuss the connection between the large $R$ charge and 
symmetry nonrestoration in the early Universe. We then 
turn to the minimal supersymmetric $SU(5)$ theory and its 
simple extensions which incorporate our scenario. 

\section{A prototype toy example}
\label{toy}

In order to illustrate our mechanism we discuss the 
simplest supersymmetric gauge model, that is the 
supersymmetric QED with coupling constant $g$ 
\cite{rs97}. The minimal spectrum consists of the chiral 
superfields $\Phi_+$ and $\Phi_-$ with gauge charges $+1$ 
and $-1$ respectively and with the most general 
renormalizable superpotential 

\begin{equation}
W=m \Phi_+\Phi_-\;.
\end{equation}

This model possesses an automatic $R$ symmetry, 

\begin{equation}
\Phi_\pm\to {\rm e}^{i\alpha}\:\Phi_\pm\;,\;
\theta\to {\rm e}^{i\alpha}\:\theta\;.
\end{equation}

Following the reference \cite{rs97} we assume that there 
is a non vanishing background density $n_R$ of the $U(1)_R$ 
charge. The effective potential at high temperature and 
high density can be computed using the usual techniques 
\cite{hw82,bbd91}:

\begin{equation}
V_{eff}(n_R,T)=g^2T^2\phi^2+{3\:n_R^2\over 5T^2+24\phi^2}\;,
\end{equation}

\noindent
where $\phi=\phi_+=\phi_-$ is easily seen to be the 
minimum (this explains the vanishing of the $D$-term in the 
potential). It is easy to see that for 

\begin{equation}
n_R>n_{R}^{crit}={5g\over 6\sqrt{2}}T^3
\end{equation}

\noindent
$\phi$ gets a nonzero vacuum expectation value (vev) 
and breaks the $U(1)$ gauge symmetry. 

This simple example illustrates perfectly the general 
situation: if the field in question carries an $R$ charge, 
for sufficiently large values of this charge, the gauge 
symmetry will be spontaneously broken even at high 
temperature. This phenomenon takes place because the charge cannot 
entirely reside in the thermal excited modes if the conserved charge 
stored in the system is larger than some critical value: the charge must 
flow into the vacuum and this is an indication that the vev of the charged 
field is non-zero. 
A natural candidate for our considerations is represented by the 
superheavy Higgs field in the adjoint representation of a GUT 
theory to which we now turn our attention.

\section{Grand Unification and gauge charges}
\label{gut} 

In what follows we shall discuss theories made on SU(N) 
groups. Of course, the minimal supersymmetric SU(5) model 
is of our primary interest, but it will turn out that we 
must go beyond it. To set the discussion and to facilitate 
the computations we now discuss SU(N) models in general. 

With the superheavy Higgs superfield $\Phi$ being in the 
adjoint representation, the superpotential takes the form

\begin{equation}
W = m {\rm Tr}\, \Phi^2 + \lambda {\rm Tr}\, \Phi^3.
\label{gutspot}
\end{equation}

For $m=0$ the theory has a $U(1)_R$ global symmetry

\begin{equation}
\Phi \to {\rm e }^{i \alpha} \Phi \quad , \quad \theta \to {\rm e}^{i 
3 \alpha/2} \theta.
\label{rsym}
\end{equation}

The trouble is that $m \neq 0$ is necessary in order 
to achieve the possibility of non vanishing vev 
$\langle \Phi \rangle \neq 0 $ at zero temperature. 
However, for temperatures $T$ larger than $m$ one 
could still hope that the rate $\Gamma_R$ at which 
$R$-violating processes take place is  slower than 
the expansion rate of the Universe, $H$.

Now, for $m\neq 0$, just as in the toy example, we could 
achieve a non vanishing vev $\langle \Phi \rangle$ for $T\gg m$ if 
the background $R$ charge of the Universe is large enough. 
Then $\langle \Phi \rangle \sim n_R^{1/3} \sim T$. By looking at the 
equation of motion of the $R$ charge density is it easy to convince 
oneself that the rate of the $R$-violating processes is as fast as 
$\Gamma_R \simeq \Lambda^2/T$ where 

\begin{equation}
\Lambda^4 \simeq m \langle \Phi(T)\rangle^3 \simeq m T^3, 
\end{equation}

\noindent
and thus

\begin{equation}
\Gamma_R \simeq \sqrt{m T}. 
\end{equation}

\noindent
For $m \sim 10^{16}$ GeV and  obviously for $T$ larger than $10^{16}$ 
GeV there is an epoch when $\Gamma_R > H \sim \sqrt{g_*} 
T^2/M_{P}$, and thus any previous $R$ charge could have 
been washed out.  However, we cannot guarantee this without the precise 
computation of the wash-out rate. More important, at temperatures of the 
order of the GUT scale thermal equilibrium is not easy to attain and all 
the phenomena, including the GUT phase transition leading to the possible 
formation of monopoles, may have taken place out-of-equilibrium.

The above theory may not work. On the other hand, it suffers 
from a serious drawback: the large 
GUT scale $m$  is put by hand.

A more complete theory should try to compute the above 
ratio, in which case $m$ should be determined 
dynamically. This philosophy fortunately cries out for a 
global $R$-symmetry.

\subsection{SUSY GUTs with a dynamical determination 
of the unification scale}

Here the philosophy is very simple. One eliminates 
the mass term from the superpotential and attempts 
to compute the ratio of the GUT and the electroweak mass 
scales dynamically. Here the results depend dramatically 
on whether N of SU(N) (where $N>4$ in realistic theories) 
is even or odd, as we describe now. 
For $m=0$ and for a diagonal $\Phi_{ij}=\phi_i\delta_{ij}$ 
which makes the D-potential vanish, 
$F_i=0$ imply $\phi_i^2=\phi_0^2$. 
Since the trace of $\Phi$ is zero, it is easy to see that 
for odd N there is only a trivial solution $\phi_0 =0$, 
and this case will be treated separately in detail for 
the physically relevant case of SU(5).

On the other hand, for N even ($N=2n$) the solution has the form 

\begin{equation}
\langle\Phi\rangle=\phi_0 \,{\rm diag}(I,-I)\;,
\label{phi}
\end{equation}

\noindent
where $\phi_0$ denotes the flat direction and I is the 
$n\times n$ unity matrix. Thus for 
$\phi_0\ne 0$ the original SU(2n) symmetry is broken 
down to SU(n)$\times$SU(n)$\times$U(1). Obviously, 
the minimal such theory which contains the standard 
model is based on SU(6) gauge group. 
This flat direction is a characteristic of 
the $R$-symmetry above and it  is lifted with 
the soft supersymmetry  breaking terms, the same 
terms that break the $R$ symmetry. As is well known, 
along the lines of ref. \cite{w81}, these soft terms 
then induce a large vev $\phi_0=M_X \simeq 10^{16} GeV$ 
through radiative corrections along the flat direction. 
Thus, this is a perfectly consistent and realistic scenario 
with a dynamical generation of the GUT scale \cite{ddrg98}.

Now, as we said in the introduction, we must make sure 
that the original SU(2n) symmetry is not restored at the 
temperatures above the GUT scale. Namely, this is the scale 
which corresponds to the usual monopole production, since 
it is at this scale that the U(1) symmetry appears first. The 
effective potential at high temperature and high density in this case is

\begin{eqnarray}
\label{VN}
V&=&{\lambda^2\over 2}\left[{\rm Tr}
\left(\Phi^2\Phi^{\dagger 2}\right)-{1\over N}
{\rm Tr}\left(\Phi^2\right)
{\rm Tr}\left(\Phi^{\dagger 2}\right)
\right]+
g^2{\rm Tr}\left(\left[\Phi,\Phi^\dagger\right]^2\right)
\nonumber\\
&+&\left[\left({N^2-4\over 16N}\right)\lambda^2+Ng^2\right]T^2
{\rm Tr}\left(\Phi\Phi^\dagger\right)+
{n_R^2/2\over 3(N^2-1)T^2/4+4{\rm Tr}(\Phi\Phi^\dagger)}
\end{eqnarray}

It is easy to see that for R-charge density $n_R$ bigger 
than the critical

\begin{equation}
n_R^{crit}={3\over 4}(N^2-1)\left[\left({N^2-4\over
32N}\right)\lambda^2+{Ng^2\over 2}\right]^{1/2}T^3
\end{equation}

\noindent
the symmetry breaking is in the same direction as at $T=0$ 
(\ref{phi}) with $\phi_0$ now given by

\begin{equation}
\phi_0^2={3\over 16}\left({N^2-1\over N}\right)
\left({n_R-n_R^{crit}\over n_R^{crit}}\right)T^2\;.
\end{equation}

Notice that the direction of the vev of $\Phi$ is fixed by 
the supersymmetric terms $V_F$ and $V_D$ (the first 
line in (\ref{VN})) and thus obviously has the same form 
$\langle\Phi\rangle=\phi_0 \,{\rm diag}(I,-I)$ as at zero 
temperature. In this case $V_F$ and $V_D$ play no role 
in determining the critical density and the magnitude $\phi_0$. 

The fact that the vacuum has the same form at all temperatures 
is a remarkable fact and it provides a solution to another 
serious problem of supersymmetric GUTs. Namely, in the usual 
minimal GUTs with degenerate minima at zero temperature and 
symmetry restoration at high temperature the preferred high $T$ 
vacuum is for $\langle\Phi\rangle=0$. Obviously, the system prefers 
to remain in this vacuum even at $T=0$. In our case no such problem 
exists. 

In other words, for $n_R>n_R^{crit}$ there is no phase transition 
whatsoever: as the Universe cools down below $T\simeq M_X$ the Higgs 
field remains in the same broken phase. Notice that 
for the monopole problem it is 
not really essential that the direction of symmetry breaking is the 
same at high and low temperature. Even if these directions were 
different the rank of the broken group would be the same since the 
adjoint representation cannot change the rank of the original group. 
Thus, there would be in any case explicit U(1) factor below and above 
the GUT scale. This is sufficient for the solution of the monopole 
problem, as we discuss at the end of this section. The crucial 
point here is that unlike in the minimal SU(5) theory at $T=0$ 
there is a flat direction but the direction is unique. It is enough 
to eliminate the $\Phi=0$ minimum at high $T$ (as in the case 
of large charge density) and the $T=0$ minimum is necessarily in 
the right direction. 

\subsection{Realistic models}

{\bf SU(5) model.}\hspace{0.5cm} 
We look first for a situation in the SU(5) 
theory in which $m$, instead of being put in by hand, 
is the vev of a singlet field $S$. An obvious attempt, 
$W = S {\rm Tr}\,\Phi^2 + {\rm Tr}\, \Phi^3$, 
does not work, for it implies $\langle \Phi \rangle =0$. 
We must go beyond the minimal model and the simplest 
extension is to postulate another adjoint superfield 
$\tilde \Phi$ with a superpotential

\begin{equation}
W = \lambda_1S{\rm Tr}\,\tilde\Phi\Phi+
\lambda_2{\rm Tr}\,\tilde\Phi\Phi^2\;. 
\end{equation}

\noindent
Obviously, at $T=0$ one of the degenerate minima is 

\begin{equation}
\langle\Phi\rangle={\lambda_1\over\lambda_2}
\langle S\rangle\:{\rm diag}(2,2,2,-3,-3).
\end{equation}

\noindent
with $\langle\tilde\Phi\rangle=0$ and $\langle S\rangle$ undetermined. 
Of course, among other minima there is also 
$\langle\Phi\rangle$ in the diagonal direction $(1,1,1,1,-4)$. 
From the toy model example of the previous section the 
reader can easily deduce what happens at high temperature. 
Needless to say, we assume again a large $R$ charge 
background density of the Universe. As our fields $\Phi$ 
and $S$ carry non vanishing R charges just as in the previous 
case for $n_R$ sufficiently large, they will have nonvanishing 
vevs even for temperatures much above $10^{16}$ GeV. The critical 
value $n_R^{c}$ can be easily computed following the previous 
calculation.

The above superpotential has two continuous U(1) $R$-symmetries:

\begin{eqnarray}
i)&&\; 
\Phi\to{\rm e}^{i\alpha}\Phi\;,\;
S\to{\rm e}^{i\alpha}S\;,\;
\tilde\Phi\to\tilde\Phi\;,\;
\theta\to{\rm e}^{i\alpha}\theta\;;\\
ii)&&\; 
\Phi\to\Phi\;,\;
S\to S\;,\;
\tilde\Phi\to{\rm e}^{i\alpha}\tilde\Phi\;,\;
\theta\to{\rm e}^{i\alpha/2}\theta\;.
\end{eqnarray}

\noindent
with corresponding charge densities $n^{(1)}_R$ and $n^{(2)}_R$. 
In what follows we shall take $n_R^{(1)}\equiv n_R\ne 0$ and 
$n_R^{(2)}=0$. Now for us it is crucial to establish the 
nonrestoration at high temperature, but the precise value 
of the vevs is not so important. It has a generic form as 
in the example SU(2n); however since in this case it gets 
to be very complicated, we will not present it. Instead, 
we shall establish the fact of symmetry breaking and give 
the critical charge density. 

The effective potential is 

\begin{equation}
V=V_F+V_D+\Delta V_T+V_n\;,
\end{equation}

\noindent
with

\begin{eqnarray}
V_F&=&{\lambda_1^2\over 2}|S|^2{\rm Tr}\,(\Phi\Phi^\dagger)+
{\lambda_1\lambda_2\over 2}\left[S{\rm Tr}\,(\Phi\Phi^{\dagger 2})+
S^*{\rm Tr}\,(\Phi^\dagger\Phi^2)\right]\nonumber\\
&+&{\lambda_2^2\over 2}
\left[{\rm Tr}\,(\Phi^2\Phi^{\dagger 2})-{1\over 5}\left|{\rm Tr}\,
\Phi^2\right|^2\right]\;,\\
V_D&=&g^2{\rm Tr}\,\left(\left[\Phi,\Phi^\dagger\right]^2\right)\;,\\
\Delta V_T&=&{T^2\over 8}\left[(\lambda_1^2+{21\over 5}\lambda_2^2
+40g^2){\rm Tr}\,(\Phi\Phi^\dagger)+12\lambda_1^2|S|^2\right]\;,\\
V_n&=&{n^2/2\over 49T^2/3+2|S|^2+4{\rm Tr}\,(\Phi\Phi^\dagger)}\;,
\end{eqnarray}

\noindent
where we already used $\langle\tilde\Phi\rangle=0$. We assume 
that $\Phi$ can be diagonalized, which minimizes $V_D$. 
The critical charge density above which the adjoint $\Phi$ gets 
a nonzero vev and so the symmetry gets broken can be calculated
straightforwardly. If $115\lambda_1^2-21\lambda_2^2-200g^2>0$ we get 

\begin{equation}
n_R^{crit}=\sqrt{3\over 2}{\lambda_1 T^3\over 30}\left[
835-63\left({\lambda_2\over\lambda_1}\right)^2-600
\left({g\over\lambda_1}\right)^2\right]\;,
\end{equation}

\noindent
while for $115\lambda_1^2-21\lambda_2^2-200g^2<0$ the 
solution is

\begin{equation}
n_R^{crit}={49\over 12}T^3\left(\lambda_1^2+{21\over 5}\lambda_2^2+
40g^2\right)^{1/2}\;.
\end{equation}

We have establish thus that the symmetry remains broken at 
temperature above the GUT scale. The question is in which of 
the two possible directions $(2,2,2,-3,-3)$ or 
$(1,1,1,1,-4)$ the symmetry breaking takes place. As far as the 
monopole problem is concerned this is of no importance, for in 
any case we will have a U(1) factor even above the critical 
temperature. However, since the tunneling from the one minimum to the 
other is too slow, one must simply assume that we start with 
the correct vacuum (this does not have to be a global minimum). 

Our model serves to illustrate the essential 
role that $R$ symmetries play, but need not be taken 
as a final theory. The crucial outcome lies in the fact that the 
GUT symmetry would not be restored. Notice that, in all the above 
we have assumed unbroken supersymmetry. When supersymmetry is 
softly broken, $U(1)_R$ gets also explicitly broken because of 
the presence of soft trilinear scalar couplings in the Lagrangian. 
Therefore, the associated net charge vanishes \cite{rs97}. However, 
this takes place at temperatures much below the GUT scale so that 
no phase transition and subsequent monopole production may occur.

Notice further that the demand of $R$ symmetry on a full 
theory implies that the light Higgs and matter 
superfields transform non trivially under it. Thus at high 
temperature their vevs should also be non vanishing 
leading to an upside-down scenario of more symmetry 
breaking in the early Universe. 

{\bf SU(6) model.}\hspace{0.5cm} As we said above, 
we can as well enlarge the gauge group. This is 
not just model building; in the minimal SU(6) GUT with the single 
adjoint representation the idea of the dynamical generation of the mass 
scale works automatically. The point is that its superpotential 
without the mass term

\begin{equation}
W=\lambda {\rm Tr}\,\Phi^3
\end{equation}

\noindent
in this case has a nontrivial solution. Namely, 
this is the situation described above for SU(2n) with 
$n=3$. Thus we can immediately write down the critical 
value for the R-charge density,

\begin{equation}
n_R^{crit}={35\over 8}\left(18g^2+\lambda^2\right)^{1/2}T^3\;,
\end{equation}

\noindent
and the vev at high density and temperature ($T\gg M_X$)

\begin{equation}
\phi_0^2={35\over 32}{n_R-n_R^{crit}\over n_R^{crit}}T^2\;.
\end{equation}

Once we have established the phenomenon of nonrestoration 
the solution to the monopole problem is almost 
automatic. The discussion proceeds along the lines of 
\cite{dms95}. First, the nonrestoration of symmetry 
eliminates the essential cause of the problem, 
which is the overproduction of monopoles \cite{p79} during the 
phase transition via the Kibble mechanism \cite{k76}. 
This of course is not sufficient to claim the solution 
for one must worry about the thermal production at the 
temperature above the GUT scale \cite{t82}. Fortunately 
this can be easily shown to be under control as in the 
case of \cite{dms95}. 

One more important comment. The reader may worry about 
the creation of monopoles even without the phase 
transition, since in any case the Higgs field is expected 
to take random values for correlations bigger that the 
horizon. Here one must resort to the idea of primordial 
inflation which presumably took place before, say at the 
Planckian scales. In such a case the whole Universe 
should have started from a causally connected region with 
the uniformly oriented Higgs field $\Phi$. 

\section{Discussion and outlook}

It has been known for a long time that a large background charge 
density in the Universe can induce symmetry breaking at temperatures 
much above the physical mass scale of the system. This, among other 
implications may have important impact on the monopole problem. Now, 
many supersymmetric models are characterized by global $R$ symmetries, 
and in fact the supersymmetric standard model posseses 
an automatic such $U(1)$ symmetry. In this letter, we have shown that 
a corresponding sufficiently large charge asymmetry will cause GUT 
symmetry breaking even above the unification scale. This then leads to 
the solution of the monopole problem. 

As in the minimal supersymmetric $SU(5)$ model there 
is not any global $R$ symmetry, at first glance this mechanism cannot 
work in this theory. However, a simple extension which provides a 
dynamical mechanism for the generation of the GUT scale naturally 
incorporates an $R$ symmetry and therefore a dynamical generation of 
the GUT scale may imply no monopole problem whatsoever. The price 
one has to pay is the doubling of the adjoint representation and 
an additional singlet field. 

The situation is far more appealing and natural in the SU(6) extension 
of the minimal grand unified theory. Remarkably enough, as long as the 
GUT scale is generated dynamically through radiative corrections there 
is automatically an $R$ symmetry. Its large nonvanishing background 
charge in the Universe guarantees the GUT symmetry to be broken at 
high temperature and furthermore in the same direction as at zero 
temperature. The absence of a phase transition not only provides the 
solution to the monopole problem but also solves the problem of 
the high-T wrong vacuum in the usual GUTs. The point there is that 
symmetry restoration chooses the vanishing vev which at $T=0$ is 
one of the degenerate minima and the system simply prefers to remain 
in that state. No such problem is encountered in the SU(6) theory with a 
large enough $R$ charge. In a sense this theory is tailor fit for the 
ideas described here. 

The remarkable 
feature of this scenario is that the Universe today is left with no 
trace of the background charge, since at temperatures below the GUT 
scale the presence of the  soft supersymmetry breaking terms will 
necessarily imply its washout.

\section*{Acknowledgements}

We are deeply grateful to Gia Dvali for his interest, crucial 
comments and advice, and invaluable discussions. 
This work outgrew of the Extended Workshop on the 
Highlights in Astroparticle Physics held at ICTP from 
October 15 to December 15, 1997. We wish to acknowledge 
useful discussions with many of the participants, 
especially Charan Aulakh and Alejandra Melfo. 
B.B. and A.R. wish to thank ICTP for hospitality. 

The work of B.B. was supported by ICTP and the  Ministry 
of Science and Technology of Slovenia. The work of 
G.S. is supported in part by EEC under the TMR contract 
ERBFMRX-CT960090.

\end{document}